\documentclass[prl,twocolumn,english,superscriptaddress]{revtex4-1}

\usepackage[T1]{fontenc}
\usepackage[latin9]{inputenc}
\usepackage{color}
\usepackage{babel}
\usepackage{latexsym}
\usepackage{float}
\usepackage{amsmath}
\usepackage{amsfonts}
\usepackage{graphicx}
\usepackage{times}   
\usepackage{esint}
\usepackage{subfigure}
\usepackage{verbatim}
\usepackage[unicode=true,pdfusetitle,
 bookmarks=false,colorlinks=true,citecolor=blue,urlcolor=blue,linkcolor=red]{hyperref}

\makeatletter

\@ifundefined{textcolor}{}
{%
 \definecolor{BLACK}{gray}{0}
 \definecolor{WHITE}{gray}{1}
 \definecolor{RED}{rgb}{1,0,0}
 \definecolor{GREEN}{rgb}{0,1,0}
 \definecolor{BLUE}{rgb}{0,0,1}
 \definecolor{CYAN}{cmyk}{1,0,0,0}
 \definecolor{MAGENTA}{cmyk}{0,1,0,0}
 \definecolor{YELLOW}{cmyk}{0,0,1,0}
}

\@ifundefined{date}{}{\date{}}
\AtBeginDocument{
  
}
\makeatother

\newcommand{\SAVE}[1]{}

\begin{document}
\renewcommand\abstractname{}
\title{Charge density waves in disordered media circumventing the Imry-Ma argument }
\author{Hitesh J. Changlani}
\affiliation{Department of Physics, University of Illinois at Urbana-Champaign, Urbana, Illinois 61801, USA}
\author{Norm M. Tubman}
\affiliation{Department of Chemistry, University of California, Berkeley, California 94720, USA}
\affiliation{Department of Physics, University of Illinois at Urbana-Champaign, Urbana, Illinois 61801, USA}
\author{Taylor L. Hughes}
\affiliation{Department of Physics, University of Illinois at Urbana-Champaign, Urbana, Illinois 61801, USA}

\date{\today}

\begin{abstract} 
Two powerful theoretical predictions, Anderson localization and the Imry-Ma argument, impose significant restrictions on the
phases of matter that can exist in the presence of even the smallest amount of disorder in one-dimensional systems. 
These predictions forbid conducting states and ordered states respectively. 
It was thus remarkable that a mechanism to circumvent Anderson localization 
 relying on the presence of {\it correlated} disorder was found, that is also 
realized in certain biomolecular systems. 
In a similar manner, we show that the Imry-Ma argument can 
be circumvented resulting in the formation of stable ordered states with discrete broken symmetries 
in disordered one dimensional systems. Specifically, we simulate a family of Hamiltonians of spinless 
fermions with correlated disorder and interactions, where we find that a charge density wave is stable up to 
a finite critical disorder strength. Having circumvented the Imry-Ma mechanism, we then investigate other 
mechanisms by which disorder can destroy an ordered state.
\end{abstract}

\maketitle

\newpage

Disorder can have drastic effects on electronic properties, especially in low dimensions. 
On the one hand, it lifts the degeneracy between competing phases through 
"order by disorder" mechanisms~\cite{Villain_order_disorder,Henley_order_disorder, Bergman_Nature07}, 
and on the other it localizes clean metallic states~\cite{Anderson1958,Giamarchi_Schulz,Belitz_Kirkpatrick,Basko06,MBL_Huse},
and even creates unusual emergent excitations~\cite{Wang_Sandvik_PRL,Changlani_percolation}. 
In one dimension, the essential physics of disorder is captured by Anderson localization~\cite{Anderson1958} 
for transport properties, and work related to the seminal paper of Imry and Ma~\cite{Imry_Ma,FishFrohSpenc,ChalkerImryMa} 
for understanding the disorder-driven destruction of ordered phases. 
The dimensionality dependence of both effects weakens with increasing spatial 
dimension, 
 thus their strongest effects are seen in one dimension.

An interesting exception to Anderson localization arises due to {\it correlations} in the disorder. In particular, Refs.
~\cite{Dunlap1990,Wu1991} discovered a class of non-interacting random $n$-mer models 
where a band of single-particle states which exhibit no backscattering exist; a condition key 
for circumventing localization. 
The focus of this article is to show that correlated disorder 
of this type can also avoid the Imry-Ma argument, and may lead to the stabilization of ordered phases in 
interacting, disordered chains.

In all $n$-mer models there are two types of `atoms', which we call $A$ and $B$, 
differing only in their on-site energy, which are placed at random on a one-dimensional chain; 
with the condition that $n$ $B$'s (an "$n$-mer") are always placed consecutively. 
With the inclusion of nearest neighbor repulsive interactions, the Hamiltonian for spinless electrons that we study is: 
\begin{equation}
	H = \sum_{i} \epsilon_i {n_{i}} -t \sum_{i} {c_i^{\dagger} c_{i+1}} + \text{H.c.} + V \sum_{i} n_i n_{i+1} 
\label{eq:H_spinless}
\end{equation}
where $c_i^{\dagger}$ ($c_i$) and $n_i$ refer to the usual spinless electron creation (destruction) and 
density operators respectively on site $i$ which is occupied by either an $A$ or $B$ atom, and correspondingly 
the on-site energy $\epsilon_i$ is either $\epsilon_A$ (set to zero throughout) or $\epsilon_B$, 
which will be referred to as the ``disorder strength," 
$t$ is the nearest-neighbor hopping parameter, and $V$ is the nearest-neighbor interaction strength. 
 We use the terminology - "monomer", "dimer", "trimer," and "quadrumer" for the cases $n=1,2,3,4$ respectively. 

Remarkably, materials with this form of disorder have been identified; for example, 
it was found that the unusual transport properties of polyaniline can be explained 
via an effective Hamiltonian approach that maps this molecule onto 
the random dimer model~\cite{Phillips1991}. Further interest in the $n-$mer models has been fueled by their possible 
relevance to describing transport in large classes of biomolecules such as 
DNA~\cite{Endres2004,Roche2003,Carpena2002,Vattay2015}. Most work on $n$-mer models has been devoted to the non-interacting case~\cite{Dunlap1990,Mondragon2013}, 
which is analytically and numerically tractable, while the interacting case has been treated 
only at the mean field level~\cite{Ordejon1994,Kosior2015}, and with exact 
diagonalization for small systems~\cite{Mondragon2014}. 
Thus, the effects of interactions on these phases are still largely unexplored. 

The disorder-free system at small $V/t$ 
is known to be a Tomonaga-Luttinger 
liquid~\cite{Giamarchi} which at 
half-filling, and a critical interaction strength ($V/t=2$), 
forms a charge density wave (CDW) state that 
remains stable for all larger $V/t$~\cite{Giamarchi,NishimotoCDW}.   
However, according to the Imry-Ma argument~\cite{Imry_Ma,Shankar1990}, 
such a state should not exist upon the slightest introduction of disorder; 
here we show how the $n-$mer models avoid this.


\begin{figure}[htpb]
\centering
\includegraphics[width=1\linewidth]{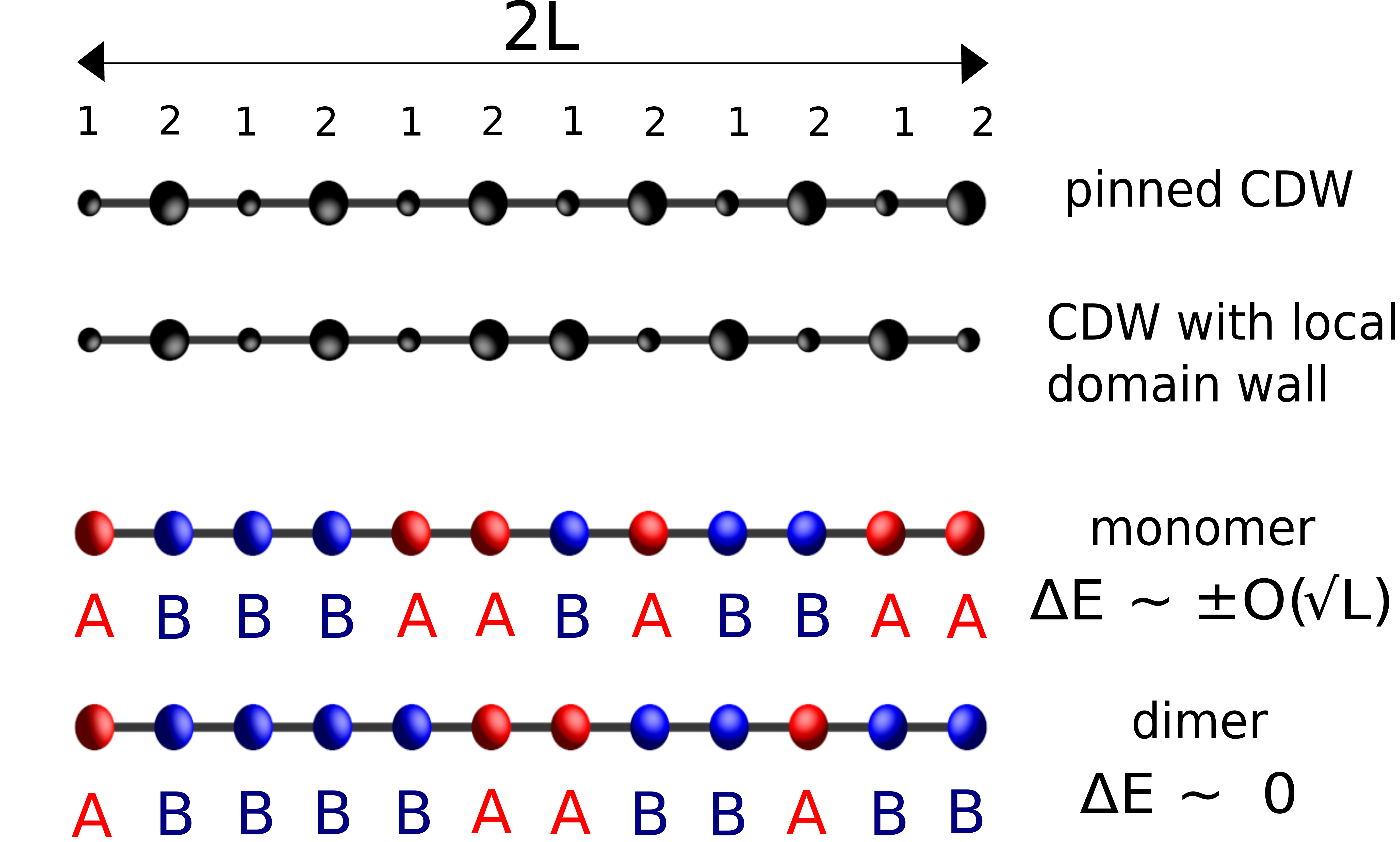}
\caption{(Color online): 
The top panel shows a schematic of the electronic density (proportional to area of circles) 
demonstrating formation of domain walls in a charge density wave in a generic disordered model. 
The Imry-Ma argument predicts that such domains are (typically) 
energetically favorable even for the smallest non-zero disorder.
The bottom panel shows schematics of a domain of length $2L$ 
in the random monomer and random dimer models. 
The red and blue sites correspond to $A$ type (monomers with on-site energy $\epsilon_A=0.0 t$) 
and $B$ type (monomers or dimers with on-site energy $\epsilon_B$) sites respectively. 
For the random monomer case, the typical difference in summed sublattice energies of 
the order of $\sqrt{L}$ while it is zero in the random dimer case.}
\label{fig:realizations} 
\end{figure}	

\begin{figure}[htpb]
\centering
\includegraphics[width=\linewidth]{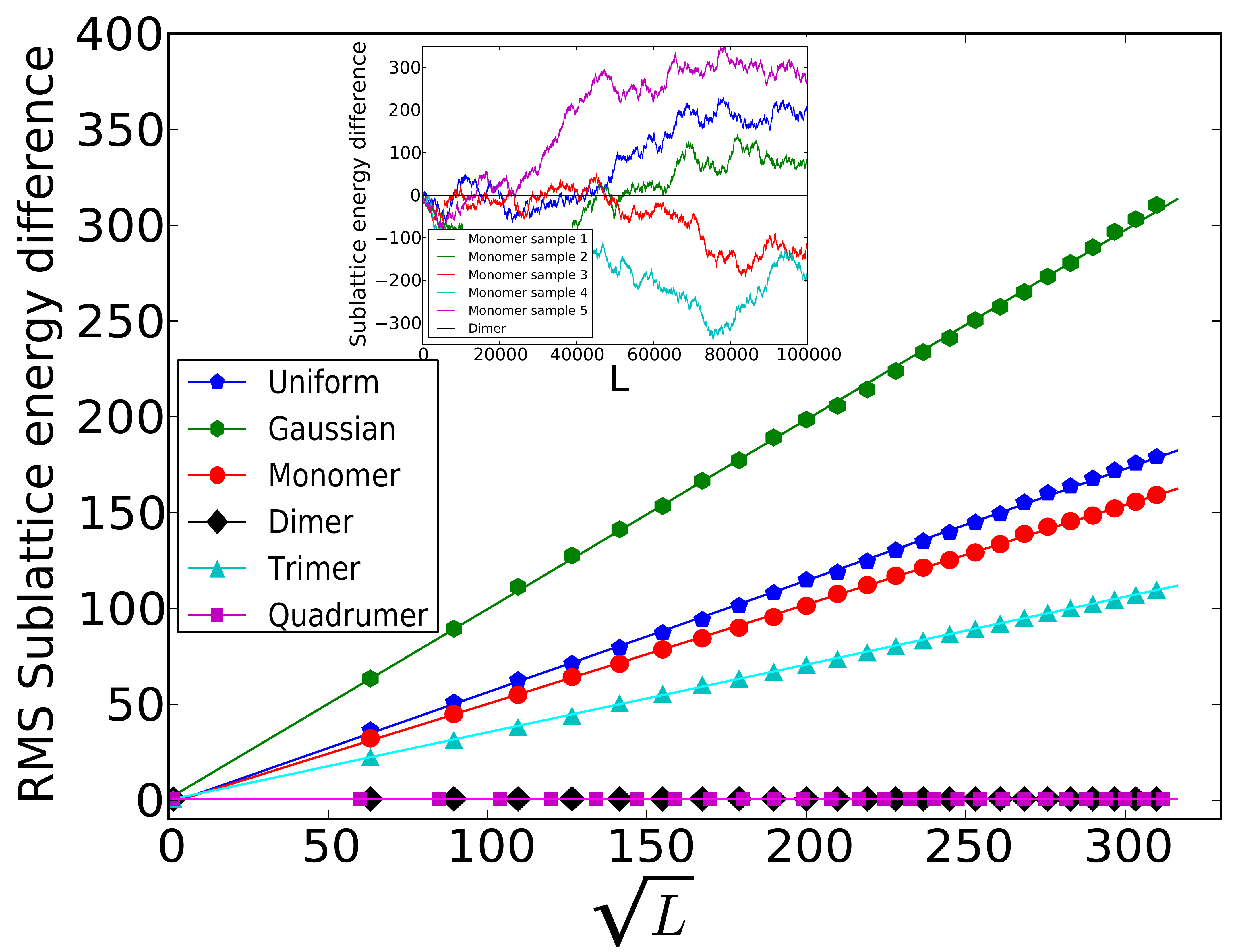}
\caption{(Color online): Root mean squared value of the difference in summed sublattice energies, 
$\Delta E,$ versus the length of the segment $L$ computed for six types of disorder distributions. 
For each disorder type, $2000$ realizations, each comprising of $10^5$ sites, were used. The uniform (box) 
distribution corresponds to maximum and minimum energies of $1$ and $-1$ respectively, the Gaussian distribution 
has a mean of $0$ and a spread ($\sigma$) of $1,$ and the monomer through quadrumer models each have
$\epsilon_A=0$ and $\epsilon_B=1$. Inset: $\Delta E$ vs $L$ for several individual disorder realizations 
of the random monomer and dimer models. The former shows large fluctuations in $\Delta E$ while 
the latter has $\Delta E=0 $ or $\pm 1$ (not visualized on the scale of the plot).}
\label{fig:disorder_sublattice} 
\end{figure}	

Given a pinned, commensurate CDW 
with every \emph{even} site (mostly) 
occupied~\footnote{We note that in any translationally invariant system with periodic boundary conditions, an exact calculation 
would yield only a linear combination of the two degenerate CDW orders; we use open boundaries, hence introducing a weak symmetry breaking due to boundary effects that will prefer occupying 
one sublattice over the other.}, as is depicted in Fig.~\ref{fig:realizations}, 
let us assess its stability to the introduction of weak disorder (for which we closely follow Ref.~\onlinecite{Shankar1990}). 
Consider a segment of length $2L$ that is part of the full 1D lattice with a large number of sites, 
and divide it into odd and even sublattices, to be labelled as 
$1$ and $2$ respectively. If the sum of all the on-site energies on the even sites is 
greater than the sum of the on-site energies on all the odd sites, then it is 
energetically favorable for 
each electron in the segment to shift by one site,  despite the cost of 
the repulsive interaction of neighboring electrons, hence forming a domain wall. 

For {\it uncorrelated} disorder, and for $L$ sufficiently large to apply statistical 
arguments, the difference between the summed energies on the two sublattices 
is of the order $\pm \sqrt{L}$. Since forming a domain wall costs only an energy of 
order $V$, the former effect always wins for some large enough $L$; hence, 
the system acts to {\it reduce} its energy by the formation of domains. 
Thus, there is no (quasi) long-range CDW order in one dimension upon 
the slightest introduction of {\it uncorrelated} disorder. 

\begin{figure*}[htpb]
\centering
\includegraphics[width=1.1\linewidth]{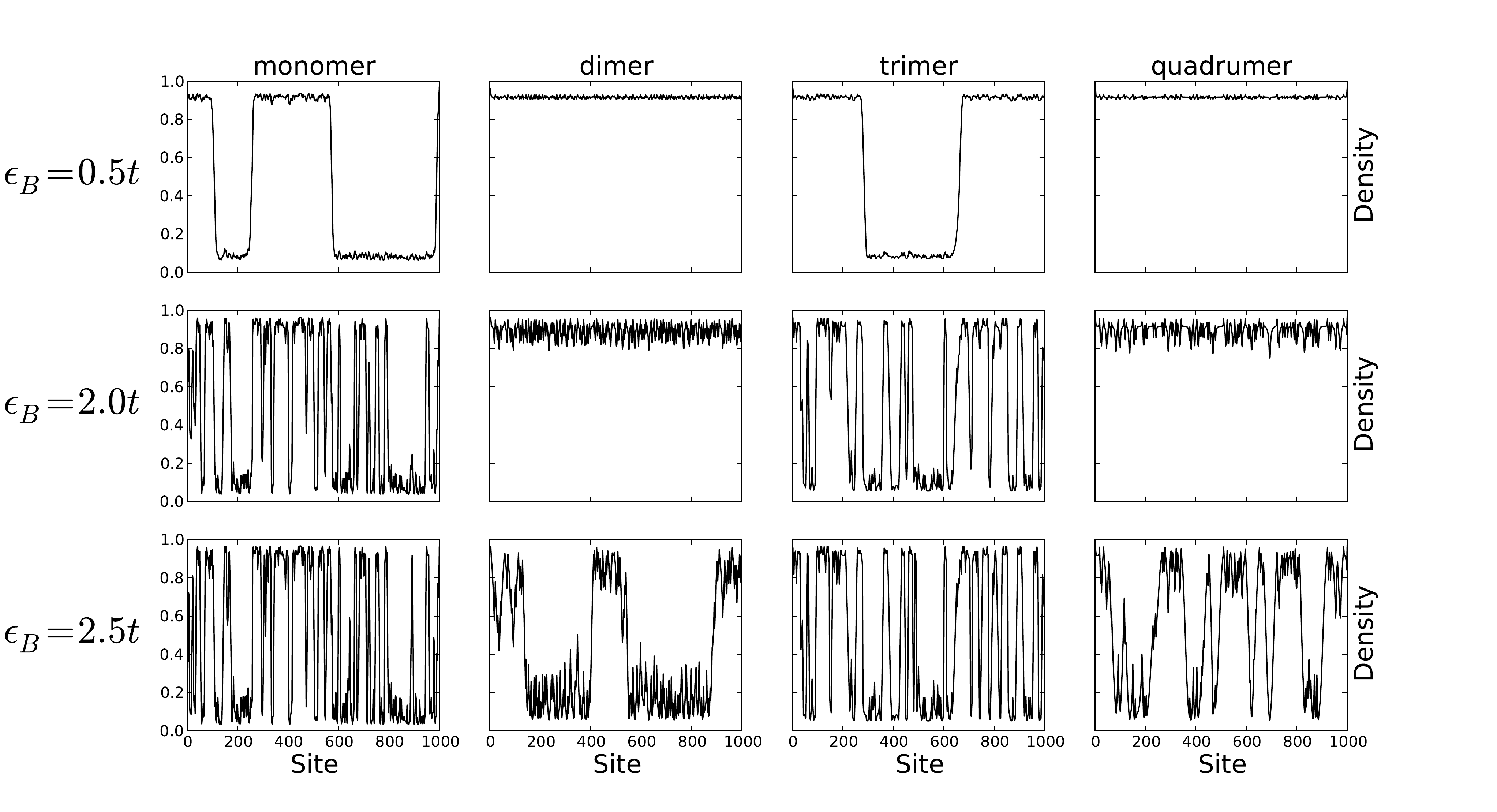}
\caption{(Color online): Fermionic density on every \emph{even} site for 
 individual realizations of the random monomer, dimer, trimer
 and quadrumer models at disorder strengths $\epsilon_B = 0.5 t, 2.0 t, 2.5 t$ for $V=5t$.
At small disorder, the odd-$n$ models show the formation of domain walls 
in agreement with the Imry-Ma argument, 
while the even-$n$ models, which circumvent the argument, 
do not (within the size considered). Beyond some critical disorder strength, 
domain wall formation is favorable for all models, i.e., 
CDW order persists only in local patches.}
\label{fig:density} 
\end{figure*}	

However, the situation is markedly different when the disorder is {\it correlated}.
Let us define $n_{\alpha,j}$ to be the number of sites where $\alpha$ is an index for the disorder site ($A$ or $B$), and $j$ 
is the sublattice index ($1$ or $2$). Then, for {\it any} disorder realization of 
the random dimer model, and \emph{any} interval of $2L$ sites, 
we have the conditions, 
\begin{eqnarray}
	n_{A,1} + n_{B,1} &=& L \nonumber \\  
	n_{A,2} + n_{B,2} &=& L \label{eq:geom_balance}. 
\end{eqnarray}
Since the instances of $B$ occur {\it only} as dimers, 
$n_{B,1}$ {\it must} be equal to $n_{B,2}$ (assuming the segment of length $2L$ does not contain any incomplete dimers) 
which, in turn, implies that $n_{A,1} = n_{A,2},$ i.e., the number of $A$-type sites on each sublattice are also 
exactly equal. 
For example, the $2L=12$ site segment of the 
random dimer disorder realization in Fig.~\ref{fig:realizations} 
has $n_{A,1}=2$ and $n_{B,1}=4$, the latter forcing the condition $n_{B,2}=4$, and hence $n_{A,2}=2$. 
 We emphasize that the relationship~\eqref{eq:geom_balance} 
holds globally and, more importantly, locally for any subset with an
even number of lattice sites. Thus, the difference between summed on-site energies on the 
even and odd sublattices is {\it zero}, i.e., 
\begin{equation}
	\Delta E \equiv \sum_{i \in 1} \epsilon_i - \sum_{i \in 2} \epsilon_i = 0. 
\label{eq:diff_sublattice}
\end{equation}
This energy difference does not grow with $L$, and therefore the Imry-Ma argument for the formation of domains is not 
expected to apply. In fact, the condition $n_{B,1}=n_{B,2},$ and hence Eq.~\eqref{eq:diff_sublattice}, 
holds for any $n$-mer model with $n$ even. This is demonstrated in Figure~\ref{fig:disorder_sublattice} 
which shows the special cancellation (or lack thereof) of the sublattice energy imbalance for the even (odd) $n-$mer models 
for an ensemble of disorder realizations. 
In instances of segments where one or more boundary cuts a dimer in half, there is 
an edge correction of one or two lattice sites, which is small on the scale of $L$ and 
does not affect our conclusions in the regime of weak-to-moderate disorder ($\epsilon_B \lesssim V $).

We verify these arguments by performing numerically accurate 
density matrix renormalization group (DMRG)~\cite{White_DMRG} calculations of the $n-$mer models for $n=1,2,3,4$, 
discussed further in the Methods section of the Supplemental Information.
Results from our simulations for individual disorder realizations are shown in Fig.~\ref{fig:density} 
where we have plotted the electronic density on every \emph{even} site 
for $V =5t$ and three disorder strengths. The boundary conditions have been chosen to 
slightly favor the high occupation of the \emph{even} sites, and thus any 
rapid decrease from high to low density is the signature of a domain wall. 

At $\epsilon_B=0.5 t,$ the random monomer and trimer models show 
large but finite domains whose size decreases with increasing disorder strength. 
In comparison, the random dimer and quadrumer show no tendency to form domain walls up to a critical 
($V$-dependent) disorder strength. For example, for all of the individual random dimer and quadrumer 
realizations in Fig.~\ref{fig:density}, the first domain walls are seen only around 
$\epsilon_B= 2.5 t$ when $V= 5t$. 

The eventual occurrence of domain walls in the random dimer and quadrumer models 
can be explained as follows. First, for sufficiently large disorder $\epsilon_B \gtrsim V$, 
the effect of the heretofore ignored edges in the $n$-mer version of the Imry-Ma argument now starts to play an important role. 
The energy of the CDW is now reduced by order $\epsilon_B$, which is greater than 
the price of forming a domain wall (order $V$). Second, 
any $B$ site would like to have lower density wherever possible, 
causing fluctuations of the density that grow large enough to destroy the ordered state. 
For example, for the realization in Fig.~\ref{fig:density}, 
for $\epsilon_B=0.5 t$, the density fluctuations are seen to be small ($ \sim 0.03$), compared to 
the maximum occupation of a site ($\sim 0.95$), and eventually grow past  
$0.5$, at which point CDW order is lost. 

This secondary mechanism can also be qualitatively understood by considering 
just a single dimer of $B$ sites placed at the center of a 1D chain 
otherwise purely made of $A$ sites. When $\epsilon_B$ is small, our numerical calculations 
indicate that the CDW is relatively robust with only a minor local rearrangement of 
electron occupations. Then, above a finite (non-zero) critical ($V$ dependent)~$\epsilon_B$, it is 
energetically favorable for the density on {\it both} the $B$ sites to be small. This creates a "phase slip" on the dimer 
forcing the rest of the chain, made solely of $A$ sites, to maintain a CDW with opposite phases 
on either side of the dimer. (Further discussions have been presented in the Supplemental Information). 
In fact, this argument holds for any even $n$-mer since favorable occupation on 
an even number of consecutive sites will cause a phase slip. 

\begin{figure}[htpb]
\centering
\includegraphics[width=0.49\linewidth]{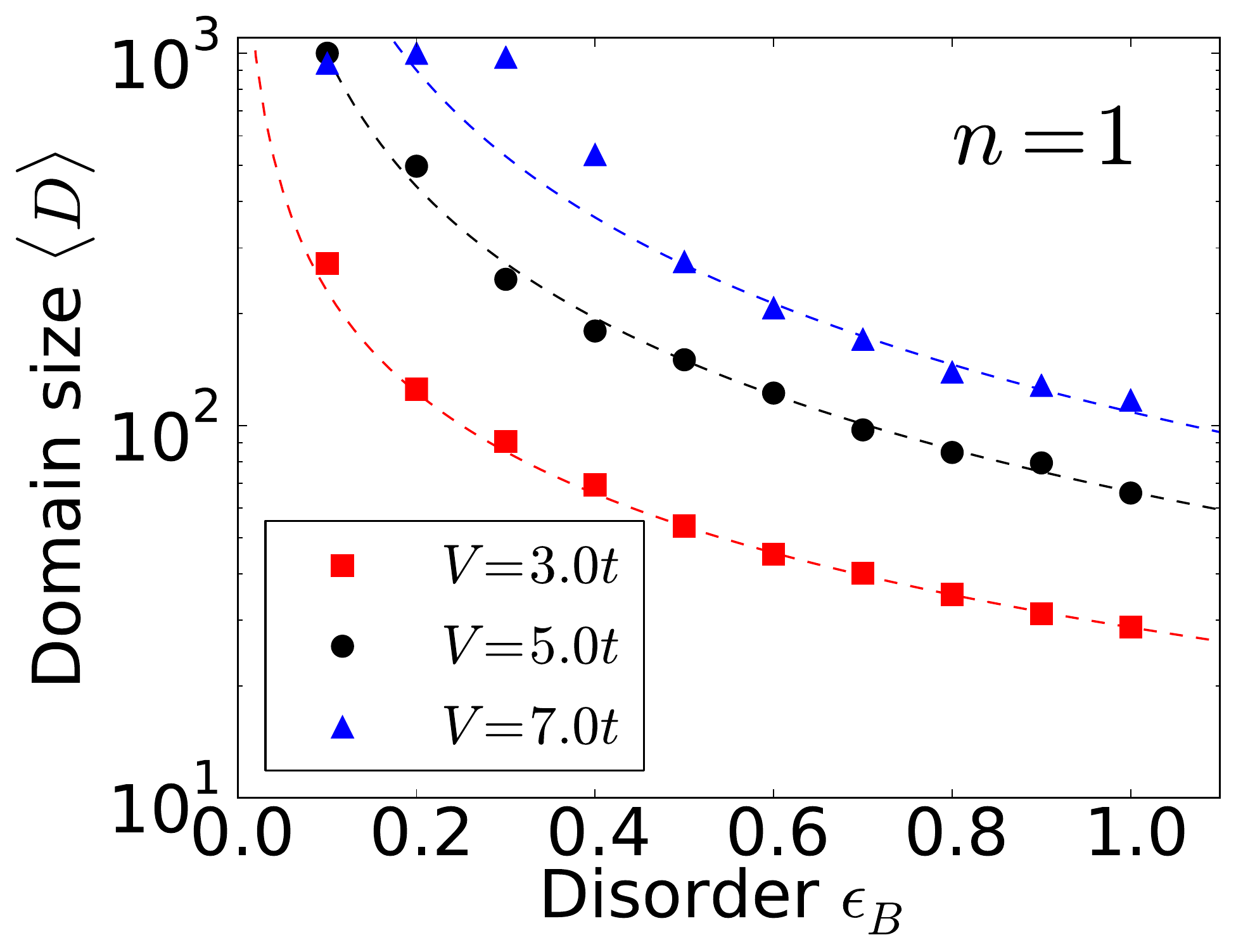}
\includegraphics[width=0.49\linewidth]{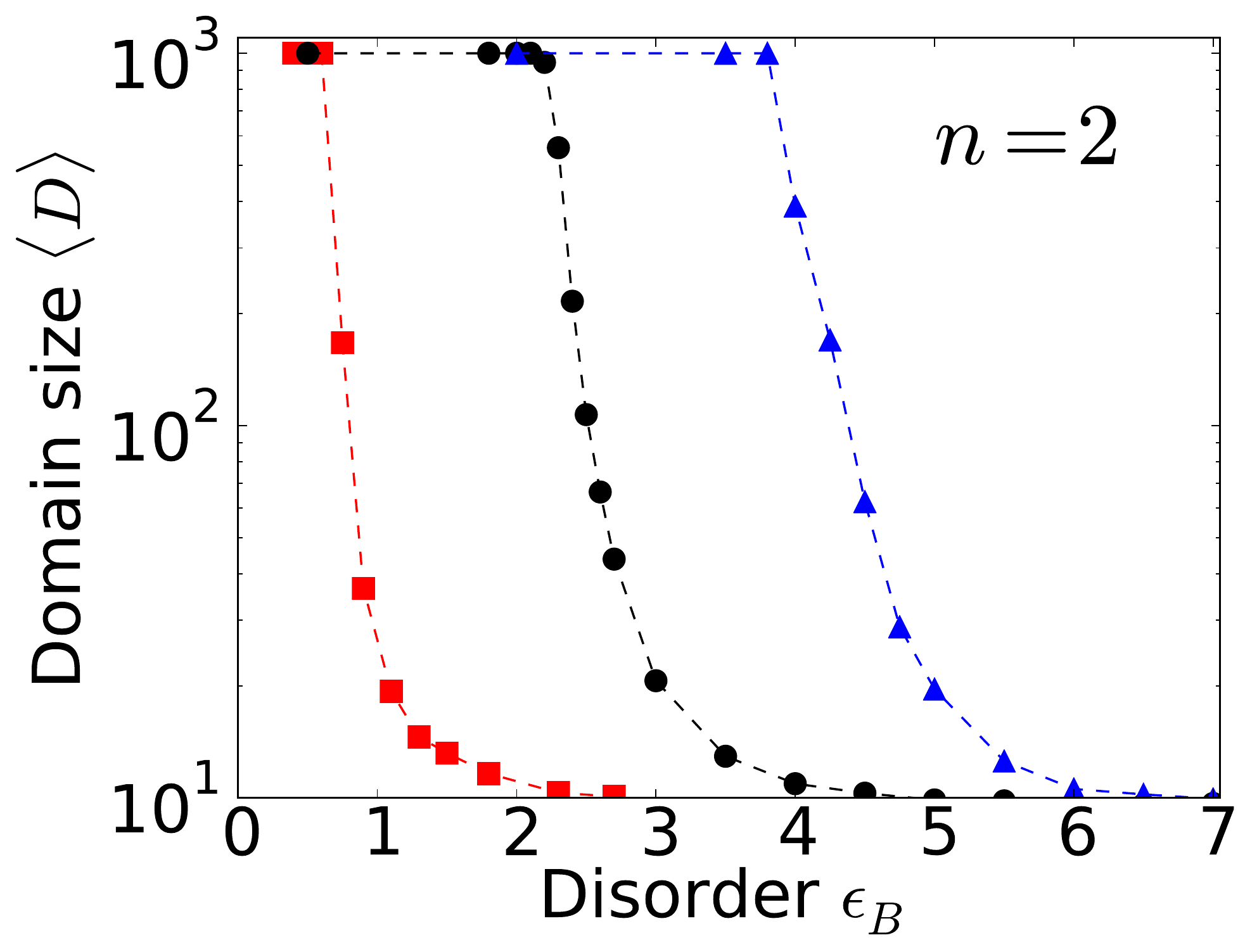}
\includegraphics[width=0.49\linewidth]{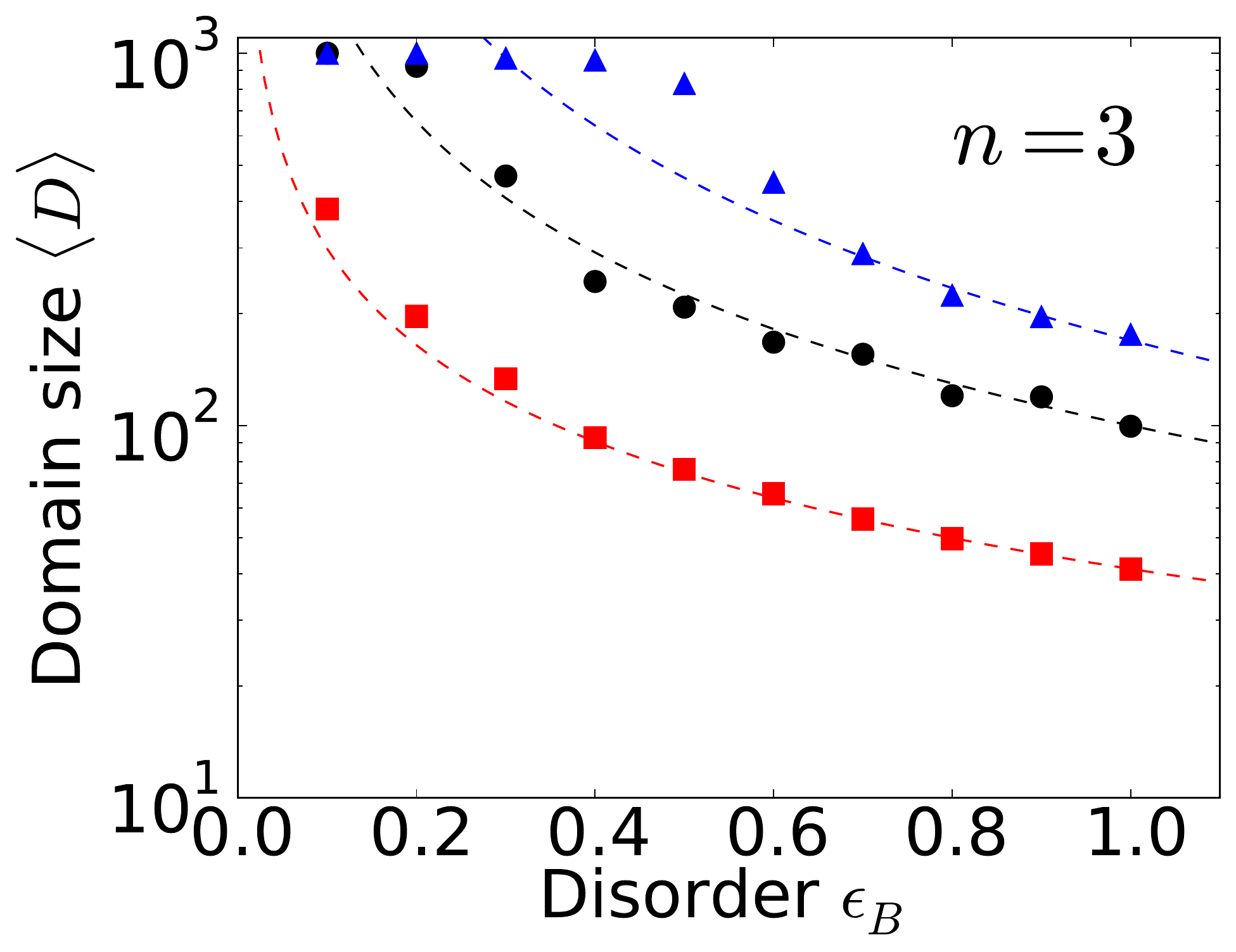}
\includegraphics[width=0.49\linewidth]{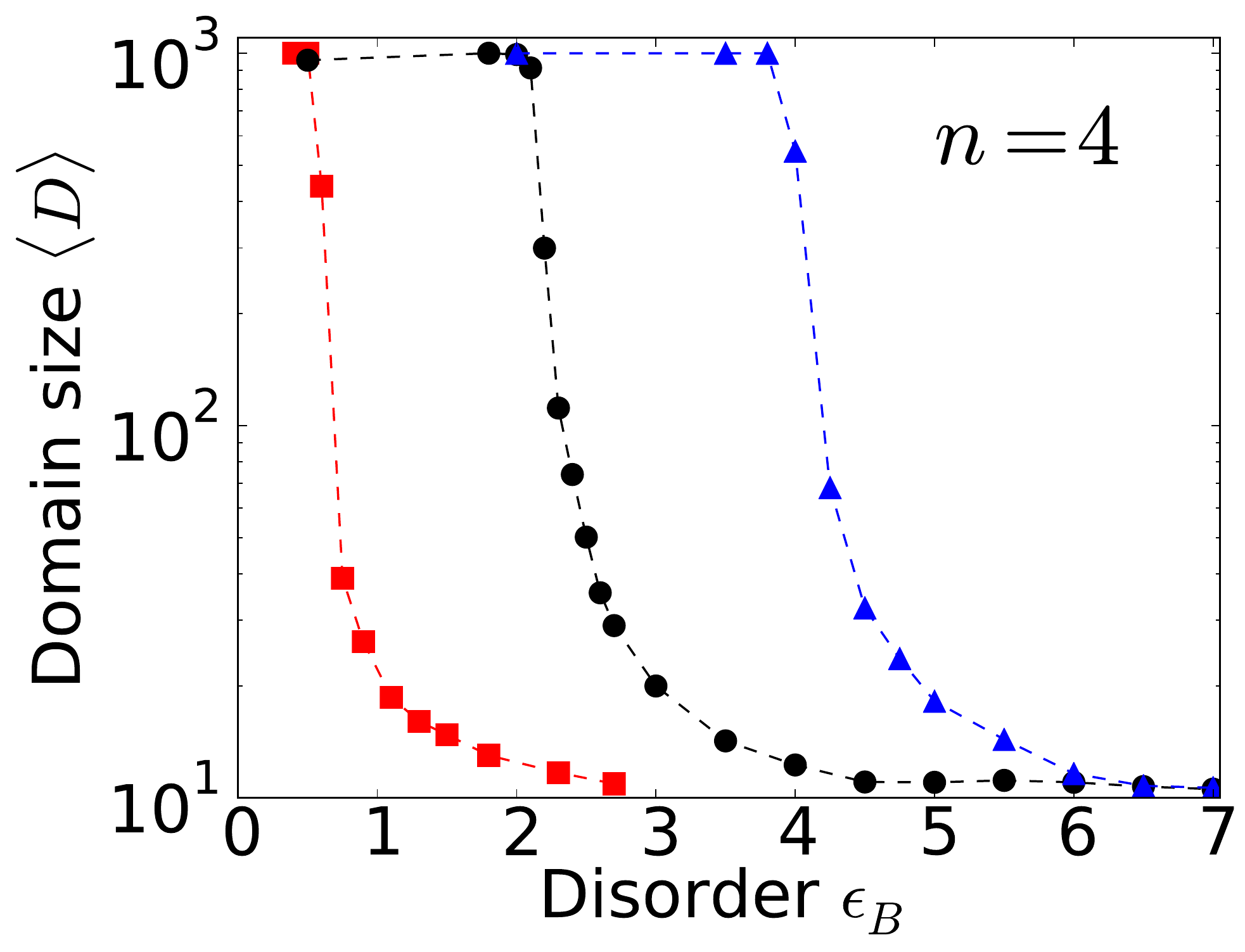}
\caption{(Color online): 
Profiles of the disorder-averaged domain size ($\langle D \rangle$) versus disorder-strength ($\epsilon_B$ in units of $t$)
for the spinless fermion random monomer ($n=1$), dimer ($n=2$), trimer ($n=3$) and quadrumer ($n=4$) 
models at half filling at various interaction strengths $V$. Around $80$ disorder realizations, 
each of $1000$ sites, were used for the averaging procedure. The dashed lines indicate approximate trends 
and serve as guides to the eye.
The critical disorder strength for the occurrence of finite domains in the odd $n-$mer cases is consistent with zero 
in concordance with the Imry-Ma argument. In the even $n-$mer case, the critical disorder strength is non-zero. 
}
\label{fig:domain} 
\end{figure}

Let us now look beyond individual realizations and perform statistical analyses 
of our samples; Fig.~\ref{fig:domain} shows the average size of the CDW domains 
as a function of disorder strength. As is anticipated from the Imry-Ma argument, the random 
monomer and trimer models show divergence in domain size 
around \emph{vanishing} disorder for all $V/t$ considered. 
This is in contrast to the random dimer and quadrumer models which have no domain walls 
until a \emph{critical} ${\epsilon_B}^{*}(V)$ is reached.

We emphasize that to prevent the formation of domain walls, 
the condition \eqref{eq:diff_sublattice} must be satisfied at all short and long length scales. 
For example, to show that the local cancellation is important, let us concoct disorder realizations of the following type. 
Take a randomly generated monomer chain of $L/2$ sites and define its "complement": 
form a realization of length $L/2$ where every $A$ type site is replaced by a $B$ type site and vice versa. 
Then place these two segments (sample and its complement) side by side to form a $L$-site chain. 
Each such disorder realization has $n_{A,1}=n_{A,2}=n_{B,1}=n_{B,2}=L/4$ and 
thus satisfies~\eqref{eq:diff_sublattice}, but only globally.
Interestingly, we find from numerics (shown in the Supplemental Information)
that domain wall formation is still favorable, and the Imry-Ma mechanism is still effective.
Despite this failure, this proposed construction raises the interesting possibility of constructing lattices from 
small blocks of length $\ell$ and their "complements" such that for sufficiently small $\ell$ 
condition~\eqref{eq:diff_sublattice} holds quasi-locally (and hence, also globally). 
This might provide another route to realizing a model where 
the Imry-Ma argument may be circumvented. 

In conclusion, we have explored an aspect of the interplay between 
interactions and disorder in one dimensional systems, an exciting 
avenue for both theory and experiments.
Using an interacting version of $n-$mer models where Anderson localization is avoided, 
we have explicitly shown that 
the Imry-Ma argument for destroying CDW order does not directly hold either, for even $n$. 
In the absence of a sub-dominant mechanism that destroys the order at 
small disorder strength, charge density waves are stabilized in media with correlated disorder. 
From the experimental viewpoint, of particular relevance are recent cold atom studies 
that have created and measured the strength of charge density waves in one dimensional geometries in the presence 
of quasiperiodic (correlated) disorder~\cite{Schreiber2015}; such a setup should 
provide the first controlled test of the existence of the phenomenon proposed here. 

We thank P. Phillips, K. Dahmen, M.N. Khan, V. Chua 
and especially I. Mondragon-Shem for discussions. We thank M. Stoudenmire for assistance 
with the ITensor software. HJC would like to dedicate this paper to late Professor Christopher L. 
Henley with whom he began his research in disordered systems. HJC was supported by the 
SciDAC grant DE-FG02-12ER46875. NMT was supported by DOE DE-NA0001789.  Computer time was provided by XSEDE, 
supported by the National Science Foundation Grant No. OCI-1053575, the Oak Ridge Leadership Computing Facility 
at the Oak Ridge National Laboratory, which is supported by the Office of Science of the U.S. Department of Energy 
under Contract No. DE-AC05-00OR22725 and the Taub campus cluster at UIUC/NCSA. 
TLH is supported by the US National Science Foundation under grant DMR 1351895-CAR.


\bibliography{refs}{}
\end{document}